\documentclass{emulateapj}
\usepackage[usenames]{color}
\usepackage{hyperref}
\hypersetup{
  pdftitle = {Star formation in the extended gaseous disk of the isolated galaxy CIG96},
  pdfkeywords = {pdf, hyperref, bookmarks},
  pdfauthor = {Daniel Espada et al.}
}

\newcommand{\kms}{\mbox{km s$^{-1}$}}

\newcommand{\hi}{\mbox{\ion{H}{1}}}

\slugcomment{Accepted version June 24, 2011}

\shorttitle{Star formation in the extended gaseous disk of the isolated galaxy CIG~96}
\shortauthors{D. Espada et al.}

\begin{document}

\title{Star formation in the extended gaseous disk of the isolated galaxy CIG~96}

\author{D. Espada \altaffilmark{1,2,3}, J. C. Mu{\~n}oz-Mateos\altaffilmark{4,5}, A. Gil de Paz\altaffilmark{4}, J. Sabater\altaffilmark{1,6}, S. Boissier\altaffilmark{7}, S. Verley\altaffilmark{8}, E. Athanassoula\altaffilmark{7}, A. Bosma\altaffilmark{7}, S. Leon\altaffilmark{9}, L. Verdes-Montenegro\altaffilmark{1}, M. Yun\altaffilmark{10} and J. Sulentic\altaffilmark{1}}

\altaffiltext{1}{Instituto de Astrof{\'i}sica de Andaluc{\'i}a, CSIC, Apdo. 3004, 18080 Granada, Spain}
\altaffiltext{2}{National Astronomical Observatory of Japan, 2-21-1 Osawa, Mitaka, Tokyo 181-8588, Japan; daniel.espada@nao.ac.jp}
\altaffiltext{3}{Harvard-Smithsonian Center for Astrophysics, 60 Garden St., Cambridge, MA 02138, USA; despada@cfa.harvard.edu.}
\altaffiltext{4}{Departamento de Astrof\'{\i}sica y CC$.$ de la Atm\'osfera, Universidad Complutense de Madrid, Avda$.$ de la Complutense, s/n, E-28040 Madrid, Spain}
\altaffiltext{5}{National Radio Astronomy Observatory, 520 Edgemont Road, Charlottesville, VA 22903-2475}
\altaffiltext{6}{Institute for Astronomy, University of Edinburgh, Edinburgh EH9 3HJ, UK}
\altaffiltext{7}{Laboratoire d'Astrophysique de Marseille, OAMP, Universit\'e Aix-Marseille \& CNRS UMR 6110, 38 rue Fr\'ed\'eric Joliot-Curie, 13388 Marseille cedex 13, France}
\altaffiltext{8}{Dept. de F\'{i}sica Te\'{o}rica y del Cosmos, Universidad de Granada, Spain}
\altaffiltext{9}{Joint ALMA Observatory/ESO, Av. Alonso de Cordova 3107, Vitacura, Santiago, Chile}
\altaffiltext{10}{Department of Astronomy, University of Massachusetts, Amherst, MA 01003, USA}

\begin{abstract} 

We study the Kennicutt-Schmidt star formation law and  efficiency in the gaseous disk of the isolated galaxy CIG\,96 (NGC\,864; \citealt{tully88}), with special emphasis on its unusually large  atomic gas (\hi) disk (r$_{\rm HI}$/r$_{25}$ = 3.5, r$_{25}$ = 1\farcm85).
We present deep GALEX near and far ultraviolet observations, used as a recent star formation tracer, and we compare them with new, high resolution (16\arcsec, or 1.6~kpc) VLA \hi\ observations. 
 The UV and \hi\ maps show good spatial correlation outside the inner 1\arcmin , where the \hi\ phase dominates over H$_2$.  { Star-forming regions} in the extended gaseous disk are mainly located along the enhanced \hi\ emission within two (relatively) symmetric giant gaseous spiral arm-like features, which emulate a \hi\ pseudo-ring at a $r$ $\simeq$ 3\arcmin . { Inside such structure,} two smaller gaseous spiral arms extend from the NE and SW of the optical disk and connect to the previously mentioned \hi\ pseudo-ring. 
Interestingly, we find that the { (atomic) Kennicutt-Schmidt power law index} systematically decreases with radius, from N $\simeq$ 3.0 $\pm$ 0.3 in the inner  disk (0\farcm8 -- 1\farcm7) to N = 1.6 $\pm$ 0.5 in the outskirts of the gaseous disk (3\farcm3 -- 4\farcm2). 
Although the star formation efficiency { (SFE), the star formation rate per unit of gas}, decreases with radius where the \hi\ component dominates as is common in galaxies, we find that there is a break of the correlation at $r$ $=$ 1.5~$r_{25}$. { At radii 1.5~$r_{25}$~$<$~$r$~$<$~3.5~$r_{25}$, mostly within the \hi\ pseudo-ring structure, there exist regions whose SFE remains nearly constant, SFE $\simeq$ 10$^{-11}$  yr$^{-1}$}.  We discuss about possible mechanisms that might be triggering the star formation in the outskirts of this galaxy, and we suggest that the constant { SFE for such large radii $r$~$>$ 2~$r_{25}$  and at such low surface densities might be a common characteristic in extended UV disk galaxies}. 
\end{abstract}

\keywords{galaxies: spirals --- galaxies: individual (NGC~864) --- galaxies: structure --- galaxies: ISM -- stars: formation}

\section{Introduction}

For three decades it has been known that about one third of galaxies show unusually extended atomic gas (\hi ) distributions \citep[e.g.,][] {bosma81,huchtmeier82}. 
It was not until recently, with the advent of the {\em Galaxy Evolution Explorer} (GALEX) \citep{martin05}, that we could observe easily star formation (SF) far beyond the optical radius of galaxies \citep{thilker05,thilker07,thilker09,gildepaz05,gildepaz07a,boissier07}. { The star-forming regions generally show a good spatial correlation with the \hi\ component (the main phase of neutral gas at large radii) in such extended \hi\ disks \citep[e.g.][]{bigiel10a}. } 

These extended Ultraviolet (XUV) disks are characterized by very blue UV-optical colors, they can reach up to four times the optical radius (as given by the $\mu_B$=25 mag/arcsec$^2$ major axis isophote), and are mainly composed of low mass stellar clusters of 10$^3$ -- 10$^6$ M$_\odot$. Recent studies have shown the presence of moderate amounts of dust and oxygen
abundances of ~Z$_\odot$/10 \citep{gildepaz07a,dong08,bresolin09}. 
\citet{thilker07} recently classified the XUV disks into Type 1 (patchy UV emission beyond the canonical SF threshold; \citealt{toomre64,martin01}) and Type 2 (blue, large compared with the size of the
galaxy in the near-infrared disks inside the SF threshold) and found XUV disks to be relatively
frequent, ~20\% and 10\% of the overall disk-galaxy population, for respectively  Types 1 and 2.

Comparing UV and \hi\ in such extended disks allows the
study of the Kennicutt-Schmidt (KS) law, or star formation rate (SFR) versus gas surface densities $\Sigma_{SFR}$ vs $\Sigma_{gas}$ \citep{schmidt59,kennicutt98}, in the extreme low-density and often low-metallicity environments of the outskirts of galaxy disks, which can elucidate the limiting conditions for gas cloud and star formation \citep{toomre64,martin01}.  { In addition, it is essential to study the relation between the availability of \hi\ in the outer disk and the depletion time scales due to SF. Such an \hi\ reservoir should play a role in the replenishment of the molecular gas content to enable future SF.} This process can thus contribute to sustain SF over cosmological times  \citep[e.g.][]{kennicutt94,bauermeister10}.

Another key question to be addressed is what are the mechanisms for triggering SF in these low-density gaseous environments.
Several mechanisms have been proposed, both of external and internal origin, including gravitational
instabilities induced by galaxy interactions (e.g. \citealt{gildepaz05}) or the impact of dark matter subhalos
or extragalactic clouds \citep{tenoriotagle81,bekki06}, turbulence
compression, supernovae, or gaseous spiral arms \citep{elmegreen06}. 
In this regard, \citet{thilker07} found that $\sim$ 75\% of the Type-1 XUV-disk objects show some kind of
optical morphology peculiarity or \hi\ evidence for interaction/merger or a minor external perturbation,  such as primordial clouds \citep{thilker09}. For instance, NGC 4262, a member of the Virgo cluster, exhibits an extended \hi\ and UV ring-like feature probably formed as a result of a past major interaction episode \citep{bettoni10}.   

 \citet{thilker07} noted that isolated galaxies might also host a Type-1 XUV-disk. Although  studying such isolated systems can provide provide information about other mechanisms producing extended disks, still no detailed study has been performed so far.
An excellent target sample is the Catalog of Isolated Galaxies \citep[CIG,][]{kara73}.
The isolation of the galaxies in the catalog has been revised within the AMIGA project (Analysis of the interstellar Medium of Isolated GAlaxies, http://amiga.iaa.es, \citealt{verdes05,verley07a,verley07b}). The imposed isolation criterion assures that the galaxies have not suffered any major interactions in a time scale of the order of 3~Gyr \citep{verdes05}. 
 We searched the GALEX Nearby Galaxy Survey (NGS, \citealt{gildepaz07a}), Deep (DIS), Medium (MIS), and All Sky Surveys  (AIS) for galaxies in the AMIGA sample. Out of the 24 galaxies in the AMIGA sample having MIS, DIS and/or NGS  data, none showed evidence of XUV emission. We found a total of 126
objects in AIS. Out of these, we identified seven XUV-disk galaxies, four of them of Type 1 (CIG~96, CIG~103, CIG~712 and CIG~812).

In this paper we examine the properties of the extended disk in the SAB(rs)c \citep{1991trcb.book.....D} galaxy CIG\,96 (NGC\,864). 
CIG\,96 has a strong and slightly curved bar, with a bar length estimated by Fourier analysis in H-band of 26\arcsec\ \citep{eskridge02,speltincx08}. 
It is located nearby { at a distance of 17.3~Mpc ($V$ = 1561.6~\kms , \citealt{espada05}, using a H$_0$=75~\kms~Mpc$^{-1}$)}, which yields a linear scale 1\arcsec\ $\simeq$ 100~pc.
We adopted for CIG~96 the major and minor axis optical diameters D$_{25}$ =  3\farcm7 and d$_{25}$ = 2\farcm6 at the isophotal level 25 mag/arcsec$^2$ in the B-band (homogenized value from LEDA, \citealt{paturel03}). Note that the diameters in NED (NASA/IPAC Extragalactic Database) are larger, D$_{25}$ = 4\farcm7 and d$_{25}$ =3\farcm5, as in the Uppsala General Catalogue of Galaxies \citep[][]{nilson73}. However the homogenized diameters given by LEDA are consistent with recent values found in the literature, such as \citet{paturel00}.
We assume throughout the paper a position angle P.A. = 23\arcdeg $\pm$ 3\arcdeg\ and an inclination $i$ =  43\arcdeg $\pm$ 2\arcdeg , values derived from modelling of the \hi\ data cube in \citet{espada05}. These are also consistent to the { homogenized values presented in LEDA},  P.A. = 24.0\arcdeg\ and $i$ = 46.7\arcdeg . 

VLA \hi\ maps with resolutions of  about $\sim$ 45\arcsec\ revealed an unusually extended { \hi\ disk} \citep{espada05}.  Besides the optical revision of the isolation \citep{verley07a}, CIG~96 is also found to be isolated from any relatively \hi-rich galaxy within the VLA primary beam ($\sim$30\arcmin). Only a diminutive galaxy with an \hi\ mass $M_{\rm HI}$ $\simeq$ 5 $\times$ 10$^6$~M$_\odot$  located at 80~kpc from CIG~96 is its nearest companion. No tidal tails or other signs of interaction were found to the reached sensitivity.
Largest  \hi\ column densities were found within a ring feature with major and minor axis of 1\farcm5 $\times$ 1\farcm0.  
More interestingly, \citet{espada05} found a pseudo-ring feature with an approximate size of 8\farcm2 $\times$ 4\farcm6. { The galaxy \hi\ disk has a kinematic asymmetry. The rotation curves in the approaching and receding sides differ, mostly as a result of a kinematically detached clump which is evident as a secondary peak in the position velocity cut at a radii $r$ $\sim$ 6\farcm5 to the SW \citep{espada05}.}

The main goal of this paper is to shed light into the local KS laws and { Star Formation Efficiency (SFE), the Star Formation Rate (SFR) per unit of gas,} of the extended gaseous component in this isolated object, using new deep GALEX observations and high resolution  ($\sim$ 15\arcsec) VLA \hi\ observations (\S~\ref{observationReduction}).  In \S~\ref{result} we describe our \hi\ and UV maps, compare the spatial location of UV relative to \hi , and study the KS law locally for different radius { ranges, as well as the local SFE as a function of radius}. In \S~\ref{model} we compare the observed $\Sigma_{gas}$ and rotation curve with that obtained from a model fitting using optical, UV and IR photometric radial profiles. Finally we provide a discussion on the main findings and our conclusions in \S~\ref{conclusion}. 

\section{Observations and data reduction} \label{observationReduction}

\subsection{VLA \hi\ Data}
\label{sub:vlahidata}
\hi\ observations were carried out with the NRAO VLA\footnote[1]{The National Radio Astronomy
  Observatory is a facility of the National Science Foundation operated under cooperative agreement by Associated Universities,
  Inc.} in its C configuration on July 23 2005, and were combined to the D configuration data published by \citet{espada05}. We used the same correlator setting as for the D configuration data,  which provided a velocity resolution of 48.8~kHz (10.4~\kms) for the 64 individual channels after Hanning smoothing. The editing and calibration of the data were done with AIPS\footnote[2]{http://www.aips.nrao.edu/cook.html}. The data were imaged using IMAGR by combining both datasets. For the cleaning process we applied a compromise between natural and uniform weighting (weighting parameter ROBUST = 0). The average of the line-free channels has been subtracted from all the individual channels. 

{ The final rms noise level achieved after 4 (D-configuration) plus 7~hours (C-configuration) per channel is $\sigma$~=~0.4\,mJy~beam$^{-1}$, with a synthesized beam of $\sim16\farcs9$ $\times$ 15\farcs6 (or 1.7 $\times$ 1.6~kpc), and a position angle P.A. = -30.10\arcdeg. 
We obtained the integrated intensity (moment 0) map from the 27 channels with \ion{H}{1} emission, to which we applied a primary beam correction (HPBW $\sim30\arcmin$ for the VLA antennas). 
The moment 0 was masked using a 3$\sigma$ (=1.2~mJy~beam$^{-1}$) clipping for each single channel. }
The { 1$\sigma$} noise level of the moment 0 map is $ 0.02$\,Jy beam$^{-1}$~\kms , calculated as the addition in quadrature of the channel's 1$\sigma$ levels.  

The excellent agreement between the single-dish total flux, S = 102.96~Jy~\kms\ using Green Bank 43m telescope (HPBW = 22\arcmin ) \citep{haynes98}, and that obtained with the VLA C+D configuration data, S = 100.3~Jy~\kms, implies that we recover most of the flux.  Thus an extended gas component is not expected to be missing in our maps.

\subsection{GALEX NUV and FUV Data} 
\label{galexuvdata}

The GALEX observatory provides both a near UV (NUV, $\lambda$ = 1771 -- 2831~\AA) and a far UV (FUV, $\lambda$ = 1344 --1786~\AA) broadband filter, with angular resolutions (FWHM) of  5\farcs3 and 4\farcs2, respectively. The field of view of the instrument is about 1.25\arcdeg, much larger than the VLA primary beam.
CIG~96 was observed in NUV for  6.385~ks  and FUV for  1.648~ks in our GALEX Guest Investigator program 065 in Cycle 5. We also combined these data with previous observations from the archive.

Foreground stars are prominent in the NUV images and we identified and flagged them easily via their UV colors (see \citealt{gildepaz07a}).
We computed and subtracted the background from both the FUV and NUV maps.
We corrected fluxes by adopting Galactic extinction as given by \citet{schlegel98}, namely E(B-V) = 0.048.

Besides correcting for the foreground galactic extinction, we must also account for radial variations in the internal extinction. The total IR (TIR) to UV ratio, TIR/UV, is a robust tracer of internal extinction, depending weakly on the dust-to-stars geometry or the extinction law \citep[see, e.g.][and references therein]{witt00,buat05}. In order to compute the TIR radial profiles, {\it Spitzer} images at 24, 70 and 160~$\micron$ would be required, but they are not available for this galaxy. Therefore, we resorted to the FUV-NUV colour (i.e., the UV slope) as an indirect tracer of dust extinction. Instead of using the classical calibrations derived for starburst galaxies \citep{calzetti94,heckman95,meurer99}, here we made use of the recipes of \citet{munoz-mateos09}, which were calibrated on a sample of normal, nearby galaxies. From the FUV-NUV colour profile we infer the TIR/FUV and TIR/NUV ratios at each radius. These ratios are then converted into extinction radial profiles in the FUV and NUV bands independently. This was done by means of the prescriptions of \citet{cortese08}, which take into account the varying contribution of young and old stars to the dust heating. The error in the extinction is dominated  by the scatter of ratios when we calibrate TIR/FUV and TIR/NUV  as a function of colour FUV-NUV, and is estimated to be of the order of 0.4~mag.

\section{Results} \label{result}

\subsection{\hi\ Component}

We present in Fig.~\ref{fig1} (lower left panel, green) the  \hi\ integrated  density distribution. 
The \hi\ extent is D$_{\rm HI}$ = 12\farcm9 at a level 0.7~M$_\odot$~pc$^{-2}$ {(1$\sigma$ noise level, \S~\ref{sub:vlahidata})}, which results in an unusually large \hi\ to optical extent ratio of D$_{\rm HI}$/D$_{\rm 25}$ $\simeq$ 3.5{. A similar ratio is that of M83, whose XUV disk properties have been previously studied in detail \citep{thilker05,thilker07,bigiel10b}.
}

The { VLA C+D} configuration \hi\ map reveals a large variety of structures that could not be clearly distinguished in the VLA D-configuration map due to the lower angular resolution. The inner ring{, with projected diameters of 1\farcm5 $\times$ 1\farcm0,} exhibits enhanced \ion{H}{1} emission along the stellar spiral arms. The outer pseudo-ring, with an extent of 8\farcm2 $\times$ 4\farcm6, seems to be connected to the inner ring to the NE and SW of the inner disk. Overall the shape of the outer pseudo-ring resembles two giant spiral arms surrounding the optical disk. 

Although the shape of the brighter features is quite symmetric, there is also a faint diffuse \hi\ emission that is
asymmetric.
The kinematically detached component in the SW that was reported in \citet{espada05} at about a radius 6\arcmin\ is seen here as a clumpy and extended component. 
A more detailed analysis of the whole \ion{H}{1} data-cube will be presented in a forthcoming paper.

\subsection{FUV and NUV Emission}
\label{sfr}

\begin{figure}
\centering
\includegraphics[width=7cm]{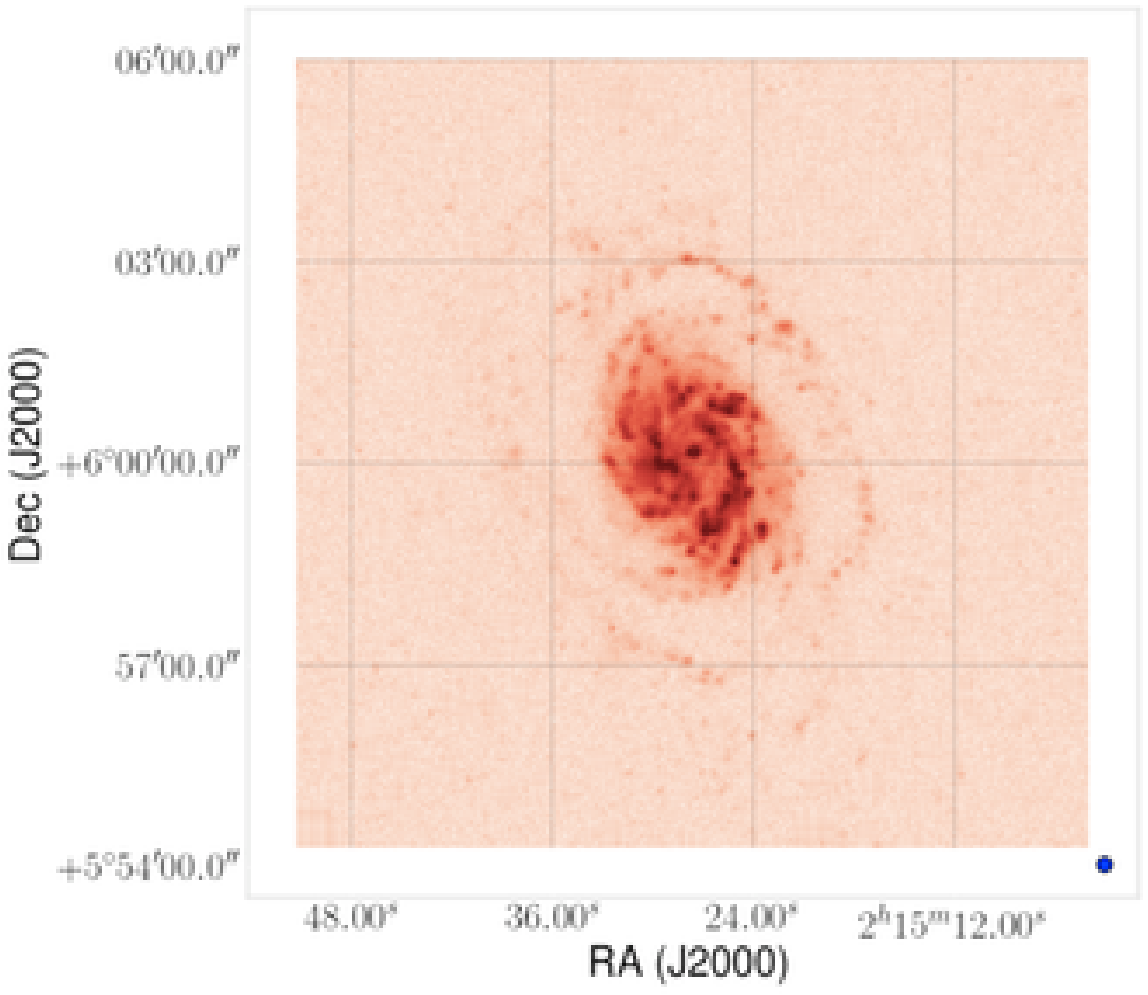}
\includegraphics[width=7cm]{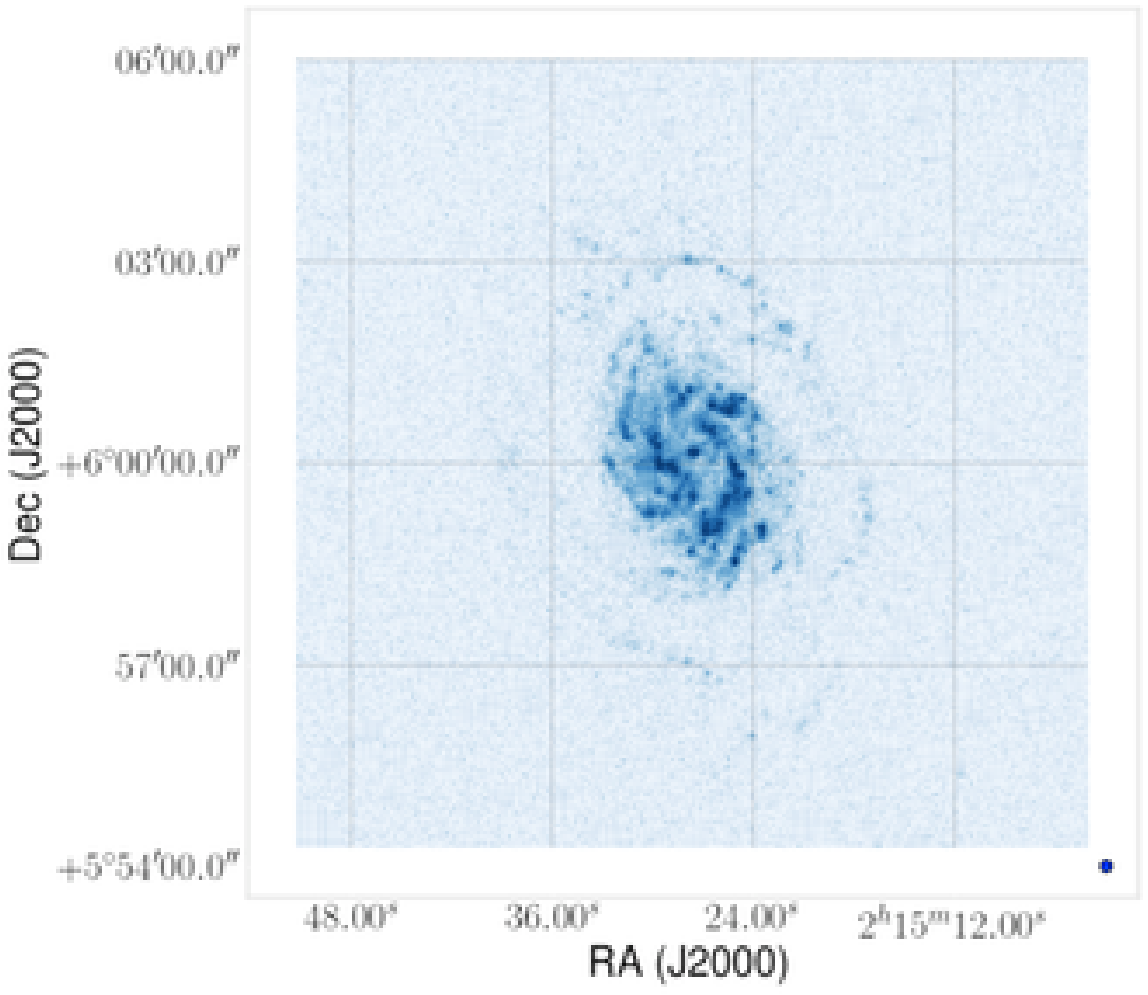}\\
\includegraphics[width=7cm]{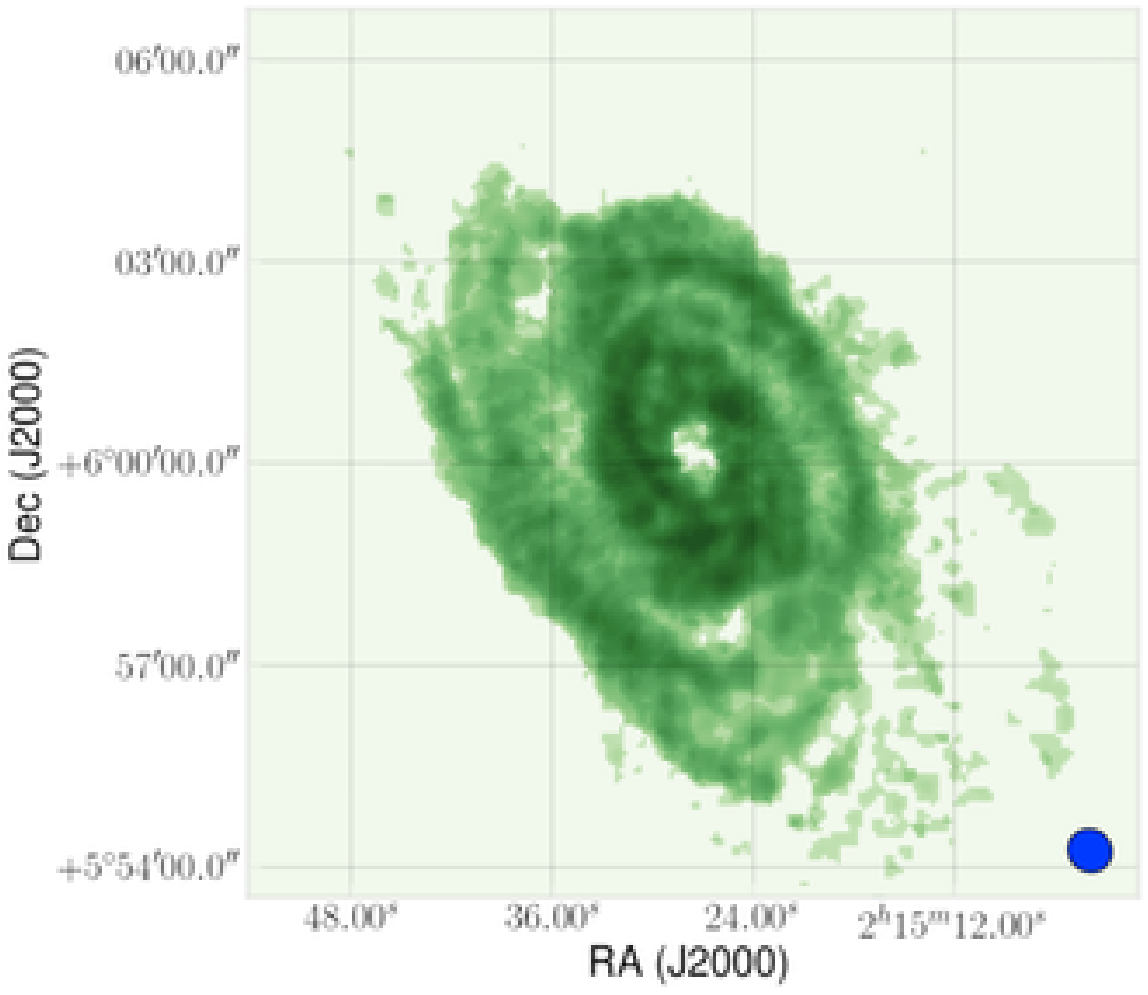}
\includegraphics[width=7cm]{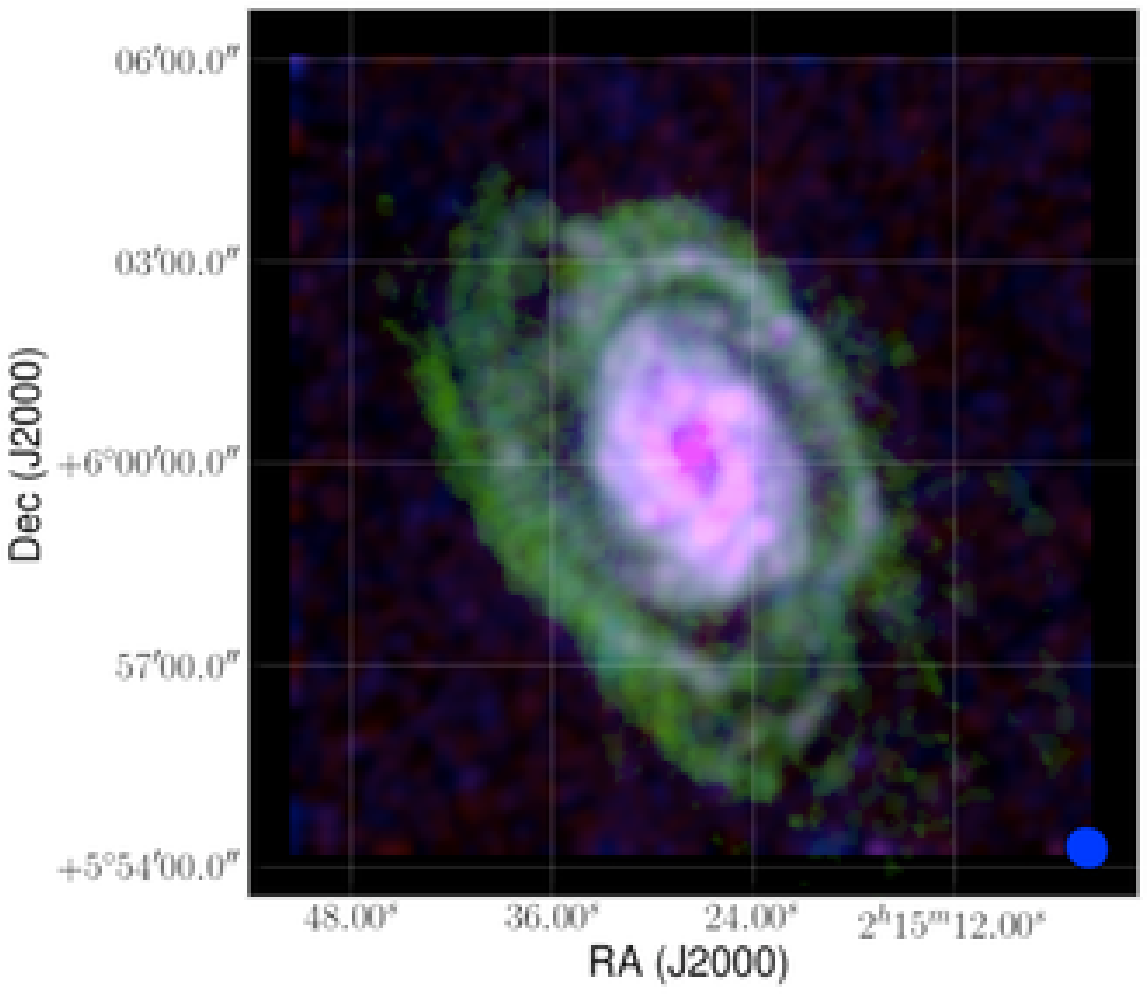}
\caption{ GALEX near UV (upper left, red) and far UV (upper right, blue) maps, as well as the VLA (C+D configuration) \hi\ moment 0 map of CIG~96 (lower left, green), and a composite of the three maps (lower right) where the resolution of the UV maps have been degraded to that of the \hi . All the panels cover a field of view of 13\arcmin $\times$ 13\arcmin. The resolution of each map is shown in the lower right. \label{fig1}}
\end{figure}

We present in Fig.~\ref{fig1} the NUV map (upper left panel, red)  and FUV map (upper right panel, blue) with the same field of view than that of the \hi\ map. Both NUV and FUV maps show an almost identical distribution. However, while the radially averaged  FUV--NUV color $\sim$ 0.5 in the internal parts of the disk  (in units of AB magnitudes per square arcsec), it reaches FUV--NUV = 0.1 -- 0.2 in the most external regions.

First, there are two nearly symmetric arms, one starting at the E of the bar and extending towards the N, and another one starting at the W of the bar and extending toward the south. 
The SW arm's appearance in both NUV and FUV emission is clumpy. 
This is similar appearance to that observed by \citet{eskridge02} in the optical and NIR.
\citeauthor{eskridge02} claim that the spiral pattern is flocculent in the outer parts of the inner optical/NIR disk ($r$ $<$ 1\farcm5). 
These flocculent regions are related to the connections between the outer \hi\ pseudo-ring  and the inner \hi\ ring to the NE and SW at the P.A. of the inner disk.
A central concentration is found unresolved, which is likely connected to a nuclear starburst \citep{martini03}. 

Second, a more diffuse and extended component of UV emission is seen in the UV maps. 
Two symmetric spiral arms are apparent, extending from the NE and SW of the optical disk. These spiral arms join an outer pseudo-ring structure to the E and W starting at about $r$ = 2\arcmin . The pseudo-ring structure is the same as that seen in the \hi\ outer disk, although the star-forming regions seem to be more confined to regions with high \hi\ surface density. 
Other filamentary and more chaotic structures are seen out to almost $r$ = 6\arcmin, especially to the SW and to a minor extent in the NE. 

\begin{figure}
\centering
\includegraphics[width=7cm]{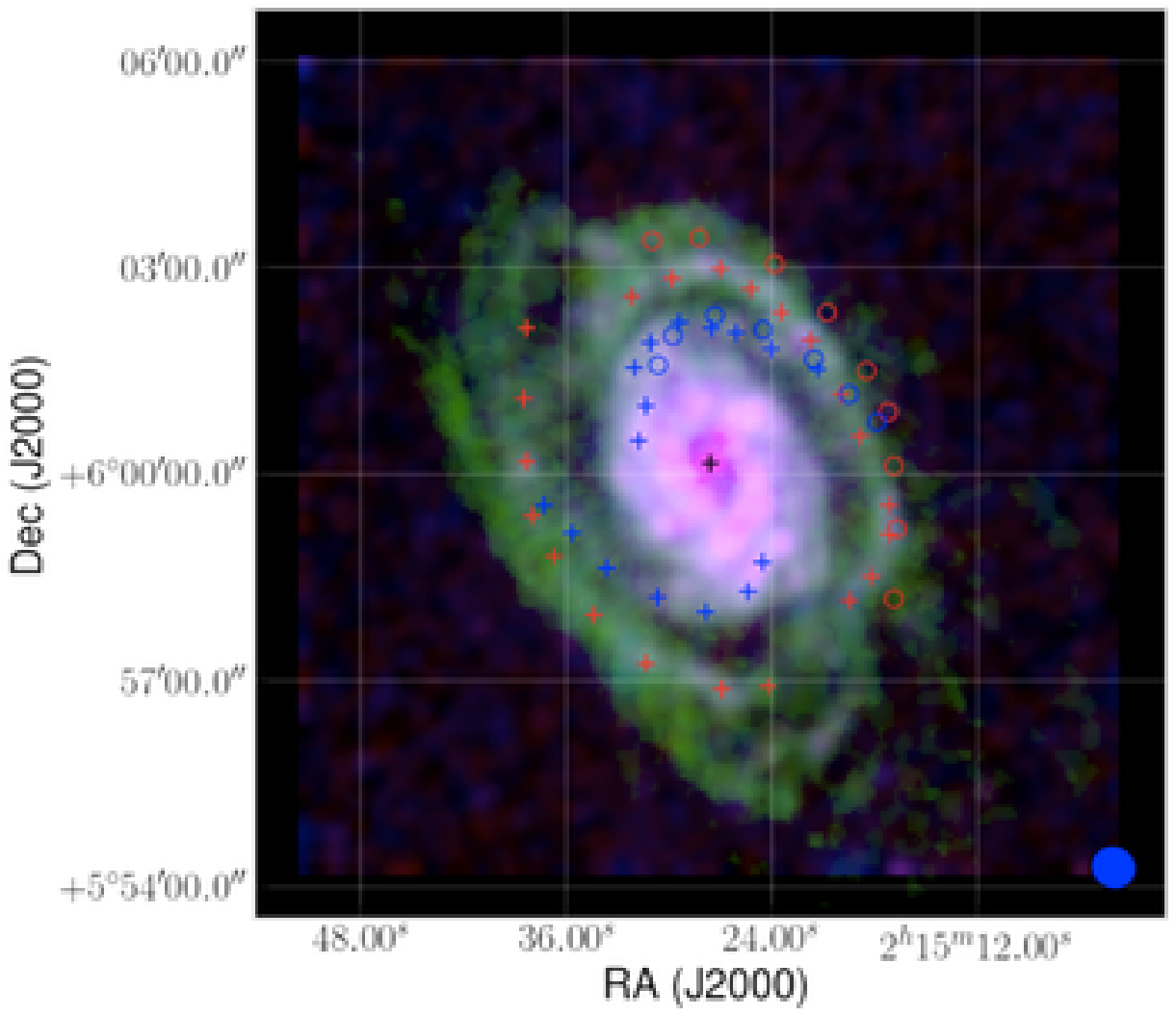}
\includegraphics[width=7cm]{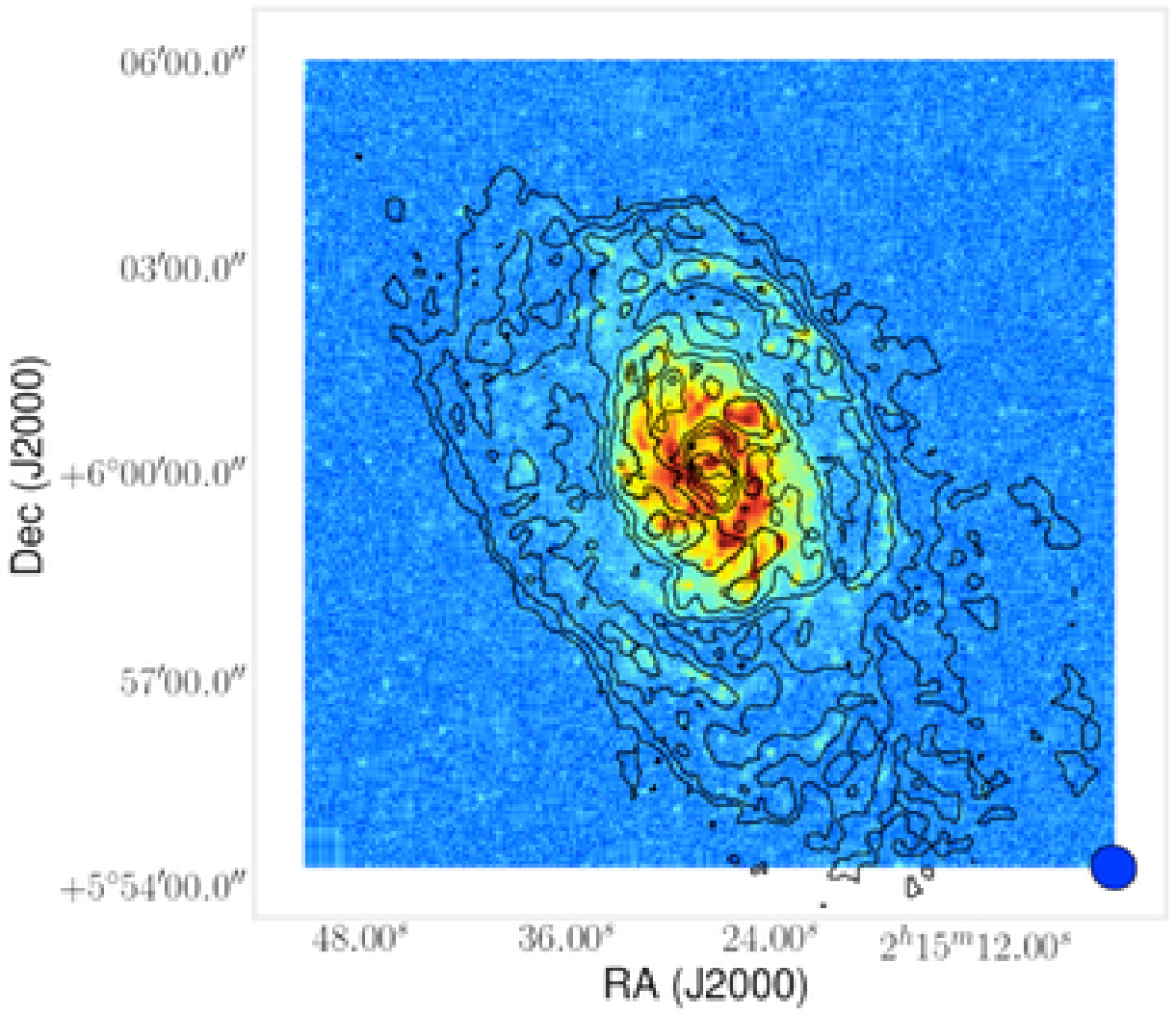}
\caption{Left panel) Scheme (over the composite image) of the gaseous spiral arms (blue cross signs) extending from the N to the W and from the S to the E, and the pseudo-ring structure (red cross signs). The circle signs indicate the SE spiral arm and the SE pseudo-ring side, rotated by 180$^{\rm o}$.  Right panel) \hi\ map in contours over the NUV color {(logarithmic) scale map. The \hi\ contours are: 0.02, 0.08, 0.16, 0.24 and 0.32 ~Jy~beam$^{-1}$~\kms }, or $\Sigma_{\hi}$ = 0.5, 1.9, 3.9, 5.8, 7.8~M$_\odot$ pc$^{-2}$.  
\label{fig2}}
\end{figure}

A composite image of the \hi\ and NUV, FUV maps convolved to the resolution of the former is presented
in Fig.~\ref{fig1}  (lower right panel).
An scheme (over the composite image) of main features in the outskirts of CIG~96 is shown in the upper panel of Fig.~\ref{fig2}: spiral arms (blue cross signs) extending from the N to the W and from the S to the E, connecting the pseudo-ring structure (red cross signs). The circle signs indicate the SE spiral arm and pseudo-ring (SE side too) rotated by 180$\rm ^o$, which shows that these features are nearly symmetrical.

The lower panel of Fig.~\ref{fig2} shows that the NUV (colour pixel map) and \hi\ maps (contour map) have a high degree of correlation. This tight spatial correlation of  \hi\ and UV emission remains far into the outer pseudo-ring, and a diffuse component all over the \hi\ envelope, as it can be seen in the last plotted contour.

\subsection{Star Formation Law} \label{SFlaws}
{ 
We obtained the KS law \citep[][]{schmidt59,kennicutt98} pixel-by-pixel by comparing the  gas surface density ($\Sigma_{gas}$) map obtained from our \hi\ data (\S~\ref{sub:vlahidata}) and the SFR surface density ($\Sigma_{\rm SFR}$) from the extinction-corrected (radially) NUV data (\S~\ref{galexuvdata}). We converted the pixel size of both maps to 16\arcsec (as the \hi\ beam size).}

In the outskirts of disks the gaseous component is usually dominated by \hi\ rather than by molecular gas, so the \hi\ content is a good estimate of the total gas content, $\Sigma_{gas}$ $\simeq$  $\Sigma_{\hi}$, especially below the saturation limit between atomic and molecular gas at $\Sigma_{gas}$ $\simeq$ 12 M$_\odot$ pc$^{-2}$ \citep{martin01,wong02,bigiel08,verley09,verley10}. 

In order to calculate the $\Sigma_{gas}$ we took into account the inclination of the galaxy and we multiplied by 1.36 to include the Helium contribution  \citep{walter08,leroy08}: 

$$
\Sigma_{gas} [\rm M_\odot \rm  pc^{-2}] =  33.6 \times \Sigma_i S_i[\rm Jy\,beam^{-1}] \Delta V  [\kms]
$$

\noindent where $S_i$ is the flux density for each channel and $\Delta \rm V$ is the channel velocity width. 
{ The 1$\sigma$ noise level for $\Sigma_{gas}$ = 0.7~M$_\odot$\,pc$^{-2}$, { using the sensitivity limit reached in our \ion{H}{1}  moment 0 map} (\S~\ref{sub:vlahidata}).}

As for the $\Sigma_{SFR}$ map we used the corrected NUV data (\S~\ref{galexuvdata}). 
In the literature, FUV is more widely used in order to derive SFR as it involves emission from young stars born in the last 100~Myr,  more similar to the H$\alpha$ ($<$ 10~Myr) than to the NUV emission ($<$ 300~Myr) \citep{verley09}. However, by using NUV, we achieve a gain in sensitivity and NUV emission is not essentially different to FUV to our resolution. { This approach can also be useful for other sources with more sensitive NUV data than FUV.}
 As for the 'hidden star formation', i.e. the SF that we cannot correct from the UV color because it is completely obscured by dust in both bands, is expected to be negligible in the outer parts of disks, since \citet{prescott08} showed that there are no such regions at radii $r$ $>$ 0.6 $r_{25}$ for SINGS galaxies \citep{kennicutt03}.

We convolved the NUV map to match the resolution of the \hi\ map. Following \citet{kennicutt98},  the $\Sigma_{SFR}$ was calculated as:

$$
\Sigma_{SFR} [\rm M_\odot\,yr^{-1}\,kpc^{-2}] =10^{-0.4 \times \mu_{NUV}[\rm AB\,mag~arcsec^{-2}] + 7.41}
 $$

\noindent { assuming solar metallicity, a \citet{salpeter55} initial mass function (IMF) and that the SFR has remained constant over the last
few 10$^8$~yr, which is the typical lifetime of stars dominating the UV emission.} In this equation  $\mu_{NUV}$ is the surface brightness (in AB magnitude units) of NUV emission. To convert to the SFR derived using \citet{kroupa01}'s IMF one should multiply our SFR by a factor 0.83.  {  The 1$\sigma$ sensitivity limit is $\Sigma_{SFR}$ = 9.2 $\times$ 10$^{-6}$~M$_\odot$\,yr$^{-1}$\,kpc$^{-2}$.} In order to avoid projection effects, we have also deprojected both the $\Sigma_{SFR}$ and the $\Sigma_{gas}$ maps. 

\begin{figure}
\centering
\includegraphics[width=8cm]{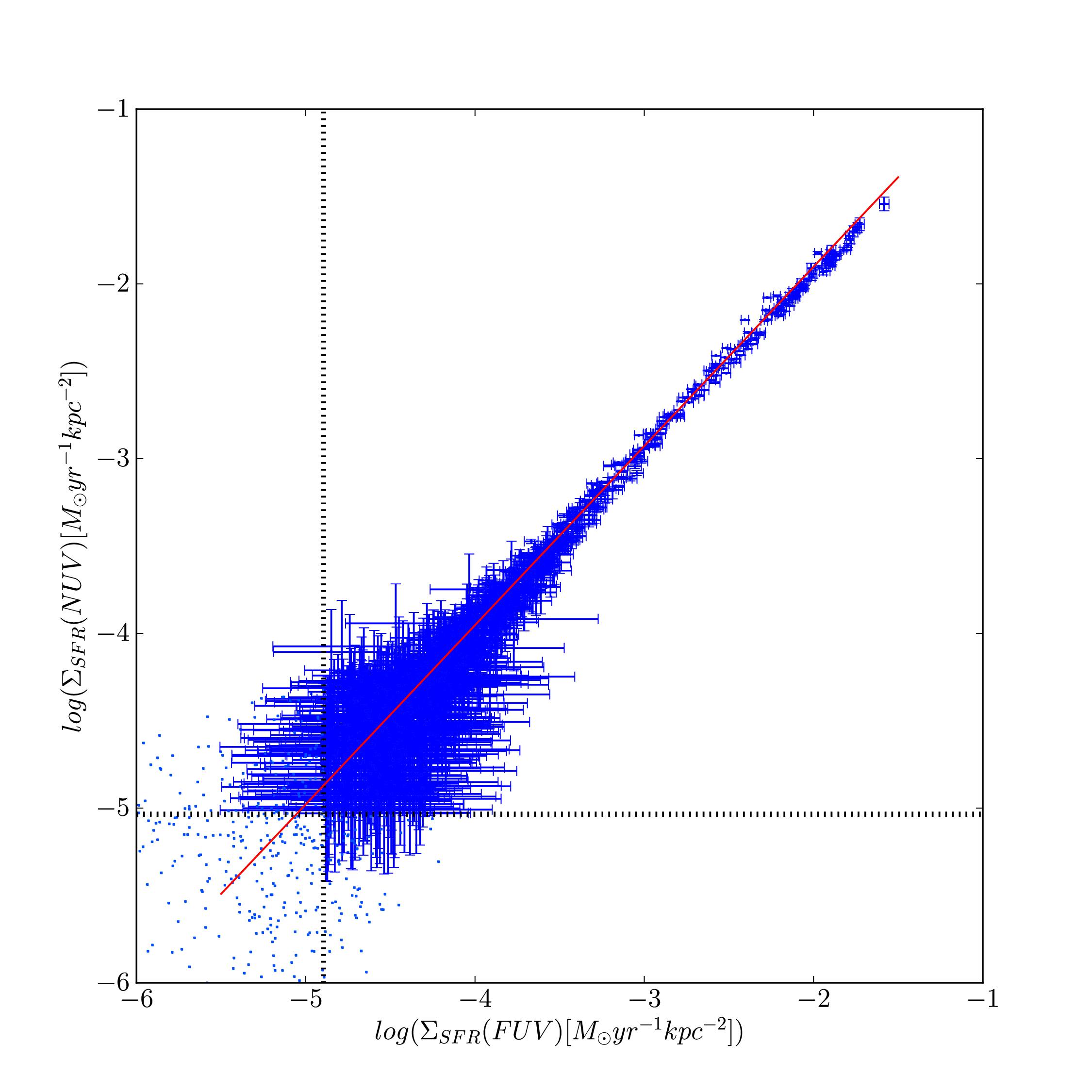}
\caption{Pixel-to-pixel (16\arcsec) comparison of SFR obtained from both extinction corrected NUV and FUV maps.  The (red) line indicates the bisector fit { using all the data points above the sensitivity limits}, with a slope equal to unity { (1.02 $\pm$ 0.03) and an intercept close to 0 (0.15 $\pm$ 0.01)}. The two dashed lines represent the noise level limit for the SFR derived from NUV (horizontal) and FUV (vertical). { Error bars have been calculated as explained in Section~\ref{SFlaws}}.
\label{figcomparanuvfuv}}
\end{figure}

{ Fig.~\ref{figcomparanuvfuv} shows the correlation between the $\Sigma_{SFR}$ derived from both the NUV and FUV maps pixels (16\arcsec size) within the inner 400\arcsec. The $\Sigma_{SFR}$ obtained from the FUV map has been calculated in a similar manner as the $\Sigma_{SFR}$  from the NUV map. The error bars presented in Fig.~\ref{figcomparanuvfuv} represent the formal uncertainties obtained from the sensitivity limit as well as from the uncertainties associated to the extinction correction. 
A bisector fit to all the data points yields a slope equal to unity and an intercept close to 0  (1.02 $\pm$ 0.03 and 0.15 $\pm$ 0.01, respectively).
We estimated the uncertainties using different random representations of the data taking into account the uncertainties (bootstrapping). We obtained similar uncertainties for the formal slope and intercept, 0.01 and 0.02, respectively.  Finally, note that calculations of $\Sigma_{SFR}$ using alternate SFR tracers give differences of the order of 50\% \citep{bigiel10a,leroy08,verley09}}. We do not consider this uncertainty in our error estimation.

\begin{figure}
\centering
\includegraphics[width=7.cm]{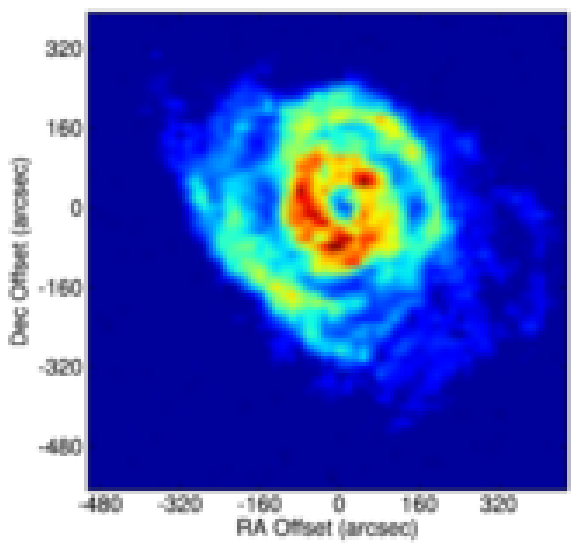}
\includegraphics[width=7.cm]{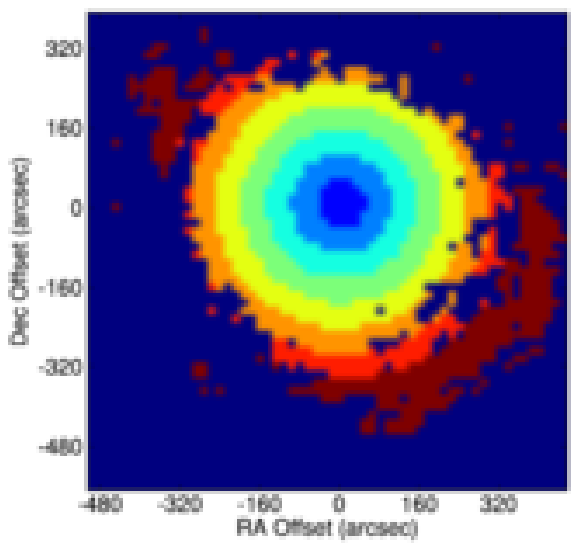}\\
\includegraphics[width=11cm]{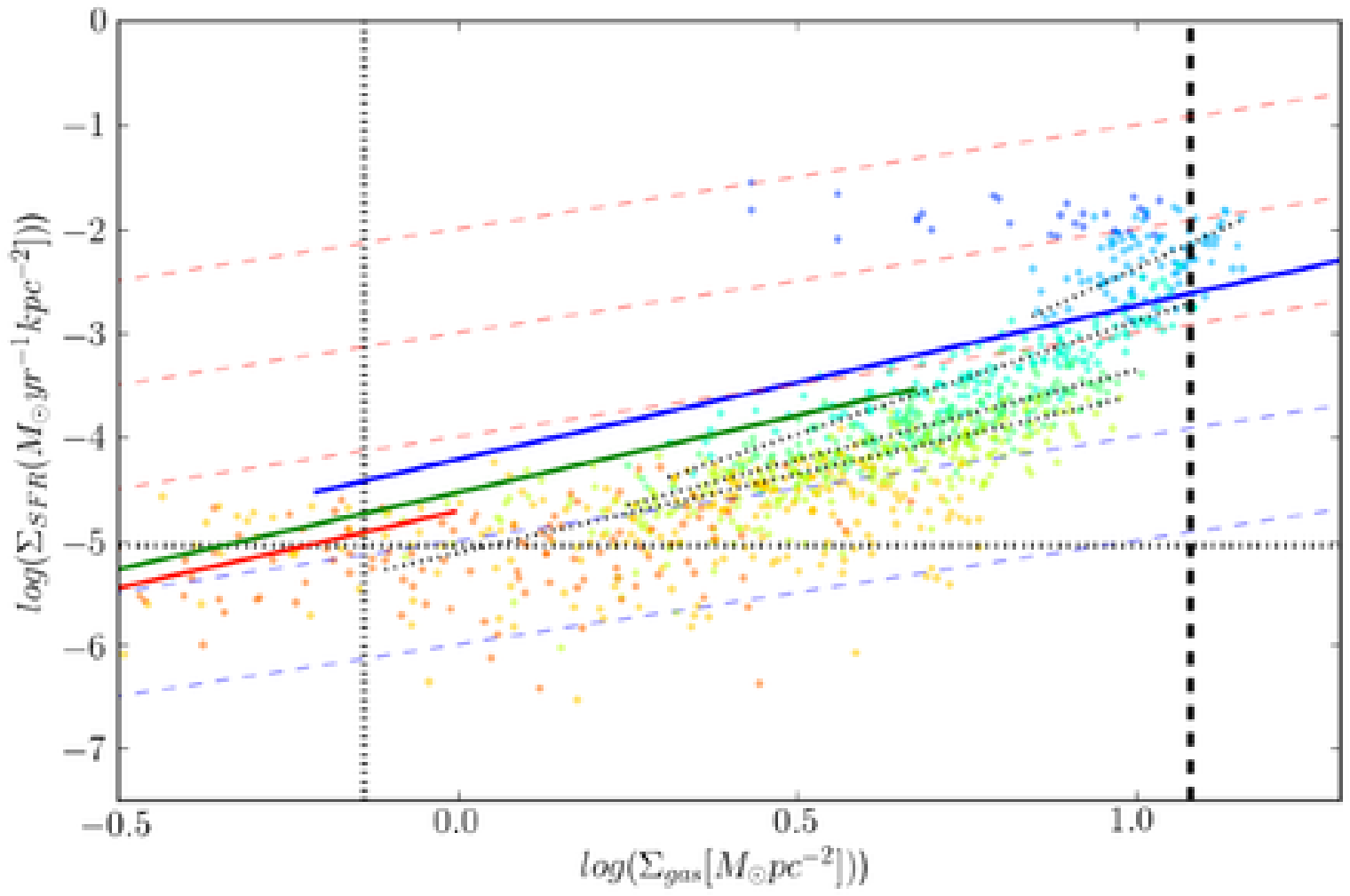}
\caption{Upper panels) Deprojected \hi\ map (left), with the colour scale showing different radial regions (right). Lower panel) $\Sigma_{\rm SFR}$ vs $\Sigma_{gas}$ (KS law) pixel to pixel for the different rings. Linear fits (bisector) to KS law for the different regions are also shown as dotted lines (See Table~\ref{tbl-1}). The diagonal (red and blue) dashed lines indicate constant SFE levels at 10$^{-12}$ (bottom line), 10$^{-11}$, 10$^{-10}$, 10$^{-9}$ and 10$^{-8}$ (upper line) yr$^{-1}$.  The vertical thick dashed line shows the threshold between molecular and atomic gas at $\Sigma_{\rm gas}$ = 12~M$_\odot$~pc$^{-2}$.  The vertical and horizontal thick dotted lines indicate the noise levels of $\Sigma_{\rm gas}$ (vertical) and $\Sigma_{\rm SFR}$ (horizontal), at 0.7\,M$_\odot$\,pc$^{-2}$ and 1.9 $\times$ 10$^{-6}$\,M$_\odot$\,yr$^{-1}$\,kpc$^{-2}$, respectively. The cross signs represent the data points in the kinematically detached region in the SW \citep{espada05}. 
\label{SFLaws}}
\end{figure}

In Fig.~\ref{SFLaws} we show the $\Sigma_{SFR}$ vs $\Sigma_{gas}$  plot for CIG~96.
From the figure we can discern the saturation limit between atomic and molecular gas found for other galaxies at $\Sigma_{gas}$ $\simeq$ 12 M$_\odot$ pc$^{-2}$. 

In order to inspect whether the local KS star formation law changes as a function of radius, we calculated a linear fit for different annuli centered at $\alpha$(2000) = 02$^{h}$ 15$^{m}$ 27\farcs64,
$\delta$(2000) =  06$^{\circ}$ 00\arcmin\ 09\farcs1 \citep{leon03}. Each colour in the data points of the  $\Sigma_{SFR}$ versus $\Sigma_{gas}$  plot in the lower panel of Fig.~\ref{SFLaws} represents a different annulus zone over the deprojected \hi\ map, as depicted in the upper panel of Fig.~\ref{SFLaws}. 

First, in the innermost region (r $<$ 0\farcm8, or 4.8~kpc; in dark blue in the left upper panel of Fig.\ref{SFLaws}), the correlation between $\Sigma_{SFR}$ versus $\Sigma_{gas}$  is biased since at such radii the contribution of  gas in molecular phase is presumably larger than that in atomic phase. For instance, M$_{\rm H2}$ = 2 $\times$ 10$^8$ M$_\odot$ using CO(1--0) observation at the Swedish-ESO 15m Submillimeter Telescope (SEST), which is characterized by a HPBW = 42\arcsec\ (radius $\sim$ 2~kpc) \citep{elfhag96,lisenfeld10}. If the molecular gas is distributed within a uniform distribution then the corresponding molecular gas surface density would be $\Sigma_{H2}$ $\sim$ 11.5 M$_\odot$ pc$^{-2}$. The $\Sigma_{H2}$ is likely larger since enhanced emission is expected toward the nucleus, inner spiral arms and along the bar. Should we include the molecular gas contribution, the correlation  $\Sigma_{SFR}$ -- $\Sigma_{gas}$ would likely continue above $\Sigma_{gas}$ $>$ 12 M$_\odot$ pc$^{-2}$ in a standard manner, with a slope N $\sim$ 1 -- 1.4 as it is observed in other galaxies \citep[e.g.][]{bigiel08}.

We find that the power index (bisector fit) of  the KS law systematically changes with radius from N = 3.0 $\pm$ 0.3 in the inner disk {(0\farcm8 -- 1\farcm7, or 4.8 -- 10.2~kpc)} to N = 1.6 $\pm$ 0.5 in the pseudo-ring feature {(3\farcm3 -- 4\farcm2, or 19.8 -- 25.2~kpc)}. Further out the correlations are biased since a considerably large amount of data points fall below the sensitivity limit. We detailed in Table~\ref{tbl-1} the parameters for the fit with $\Sigma_{gas}$ as independent variable (OLS(Y|X)) as well as the least square bisector fit, and for annuli from 0\farcm8 -- 4\farcm2 in bins of 0\farcm8 { (4.8 -- 25.2~kpc in bins of 4.8~kpc)}. {
The estimated errors of the fit parameters have been calculated via bootstrapping, in the same manner as in the $\Sigma_{SFR}$ comparison from NUV and FUV.
}  
Note that in general these parameters are not strongly affected due to spatial resolution. \citet{bigiel08} shows that the power law parameters vary only weakly with changing spatial resolution.

Besides the uncertainty that results from the assumption $\Sigma_{gas}$ $\simeq$  $\Sigma_{\hi}$, note that there is another source of uncertainty regarding a possible metallicity gradient that might affect the calculated $\Sigma_{SFR}$ (in the sense that metallicity is lower in the outskirts).
Assuming a constant calibration for a given metallicity, it can be inferred that regions with a lower metallicity would have a lower $\Sigma_{SFR}$ for a fixed UV surface brightness \citep{leitherer99}.

\subsection{Star Formation Efficiency}
\label{subsfe}

\begin{figure}
\centering
\includegraphics[width=10cm]{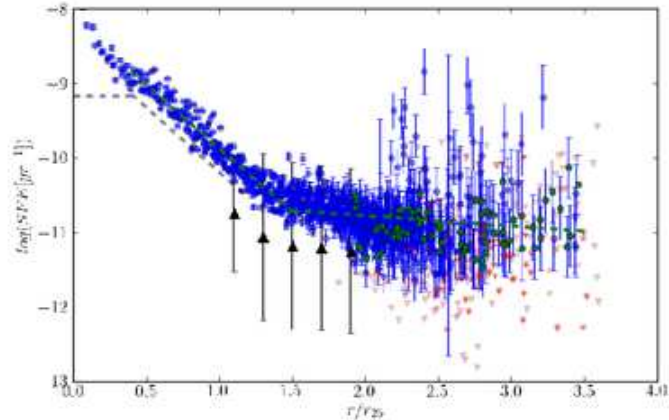}
\caption{Star Formation Efficiency (SFE) (in logarithmic scale) as a function of galactocentric radius normalized by $r_{25}$, up to $r$/$r_{25}$ = 3.5. Blue data points indicate regions whose $\Sigma_{\rm gas}$ and $\Sigma_{\rm SFR}$ both are above their noise levels, as in Fig.~\ref{SFLaws}.  There is an apparent break in the SFE fit at around $r$/$r_{25}$ $=$ 1.5. { The green dashed lines represents a fit to the detected data points: 
intercept = -8.2 $\pm$0.02, slope = -1.69 $\pm$ 0.02 if $r/r_{25}$ $<$ 1.5 and 
intercept = -10.5 $\pm$  0.1, slope = -0.14 $\pm$ 0.05 if $r/r_{25}$ $\geq$ 1.5}.  The grey dashed lines show the fit obtained by \citet{leroy08} for 18 nearby galaxies (see \S~\ref{subsfe}). Below $r$/$r_{25}$ $<$0.4 the gas surface density is likely dominated by molecular gas and the correlation to the fit does not hold. Red data points correspond to regions whose $\Sigma_{\rm gas}$ is above the noise level but $\Sigma_{\rm SFR}$ is an upper limit. Green circles represent the detected data points in the kinematically detached region in the SW \citep{espada05}. { The black triangle signs represent the median data points from \citet{bigiel10a} for a sample of spiral galaxies.} \label{sfe}}
\end{figure}

The { SFE}, the SF per unit gas mass, decreases with radius from values of 10$^{-10}$ yr$^{-1}$ in the inner ring to $\sim$ 10$^{-11}$ yr$^{-1}$ in the outskirts. In Fig.~\ref{sfe} we show the SFE as a function of radius (normalized to $r_{25}$=$D_{25}$/2). An increase in the observed scatter is present as a function of radius, from  $\sim$ 0.2  dex to $\sim$ 1 dex.
 
For comparison we have plotted in Fig.~\ref{sfe} the derived fit for the SFE as a function of radius from \citet{leroy08}:

$$ \rm SFE = 4.3 \times 10^{-10}~yr^{-1}, if~r< 0.4~r_{25} $$
$$ \rm  SFE = 2.2 \times 10^{-9}~exp(-r/ (0.25 r_{25}))~yr^{-1}, if~r> 0.4~r_{25} $$

\noindent Note that their calibration is valid up to 1.2 $r_{25}$ \citep{leroy08}.

The agreement between this fit from \citet{leroy08} and our SFE radial profile is reasonably good at 0.4~r$_{25}$~$<$~$r$~$<$~1.5~r$_{25}$. { Although our fit is slightly offset towards higher values, the slope is essentially the same in this radius range. The fit to our data and that of \citet{leroy08} are shown in Fig.~\ref{sfe} as dashed lines. 
 
 From $r$ $>$ 1.5~r$_{25}$ the fit by \citet{leroy08} would start to deviate considerably from our data, in the same manner as pointed out in other spiral galaxies by \citet{bigiel10a}. It seems to be more constant, at least for those regions where $\Sigma_{\rm SFR}$ and $\Sigma_{gas}$  are above the noise level  (blue data points). For such regions the SFE $\simeq$ 10$^{-11}$. 
  The fit to our data is shown in Fig.~\ref{sfe} as a green dashed line. It fits relatively well (rms $\sim$ 0.5 dex)  over a radius range 1.5~r$_{25}$~$<$~$r$~$<$~3.5~r$_{25}$. Note that the scatter is large enough to consider the fit parameters just as estimates. This region corresponds to the pseudo-ring feature. Many of the detected regions in the radius range 1.7~r$_{25}$ $<$  $r$ $<$ 3.5~r$_{25}$ are within the kinematically detached clump located to the SW \citep{espada05} (green circles).
  
Red triangle signs in Fig.~\ref{sfe} correspond to upper limits in $\Sigma_{SFR}$, although detected in $\Sigma_{gas}$, and thus upper limits in 
SFE. These data points would probably increase the scatter of SFE at large radii.}

\section{Model of the Multi-wavelength Light Distribution of CIG~96}
\label{model}

We inspect in this section whether the photometric properties, including the XUV emission,
is common of a spiral galaxy disk (with a given spin and mass), or whether on 
the contrary it is a peculiar object not only for its photometric properties but also for its SF history.

\subsection{Model}

 We derived photometric radial profiles for CIG~96 not only in the GALEX FUV and NUV bands, but also in the Sloan Digital Sky Survey (SDSS) \emph{ugriz} bands and the {\it Spitzer} \citep{werner04} IRAC 3.6 and 4.5 $\micron$ bands  \citep{fazio04}  \footnote[3]{We used the Basic Calibrated Data publicy available in the Spitzer archive. These images were obtained as part of the Spitzer Survey of Stellar Structure in Galaxies (P.I. K. Sheth).}.
 
 These multi-wavelength radial profiles were fitted with the models of \citet{boissier99,boissier00} with the IMF from \citet{kroupa01}, which are able to predict the radial variation of several galactic properties as a funcion of two parameters: the spin $\lambda$ (non-dimensional specific angular momentum) and the asymptotic velocity of the rotation curve $V_C$. Galaxies are modeled as a set of independently evolving rings, with the gas infall time-scale depending on both the total galaxy mass and the local mass surface density. Gas is then converted into stars following a Kennicutt-Schmidt law, modulated by a dynamical term that mimics the periodic passage of spiral arms (see also \citealt{boissier03a}). Newly-born stars follow a user-specified IMF (here we rely on that of \citealt{kroupa01}). Stars of different masses die at different rates, enriching the ISM with metals that are incorporated in subsequent generations of stars. The lifetimes, yields, evolutionary paths and spectra of stars at different radii are computed as a function of the local metallicity. An initial model was first calibrated to reproduce observables in the Milky Way \citep{boissier99}. This model was then generalized to other galaxies \citep{boissier00}, using the scaling laws derived from the $\Lambda$-Cold Dark Matter scenario \citep{mo98}.

The GALEX, SDSS and Spitzer photometric profiles were simultaneously fitted with these disk evolution models. The internal extinction profiles in the optical and near-IR bands were derived from the UV ones using a MW extinction law and a sandwich dust-to-stars geometry (further details on the fitting procedure can be found in \citealt{munozmateos11}). By minimizing the $\chi^2$ between the observed and predicted photometric profiles, the best-fitting values were found to be $\lambda = 0.048^{+0.013}_{0.012}$ and $V_C = 161^{+12}_{-9}$\,km s$^{-1}$. Using other models characterized by an IMF with less massive stars (i.e. \citealt{kroupa93}) or excluding the XUV disk further out than 150\arcsec\ does not successfully reproduce both photometric profiles and the observed \hi\ rotation curve as the ones used here do \citep{espada05}. 

\begin{figure}
\centering
\includegraphics[width=12cm]{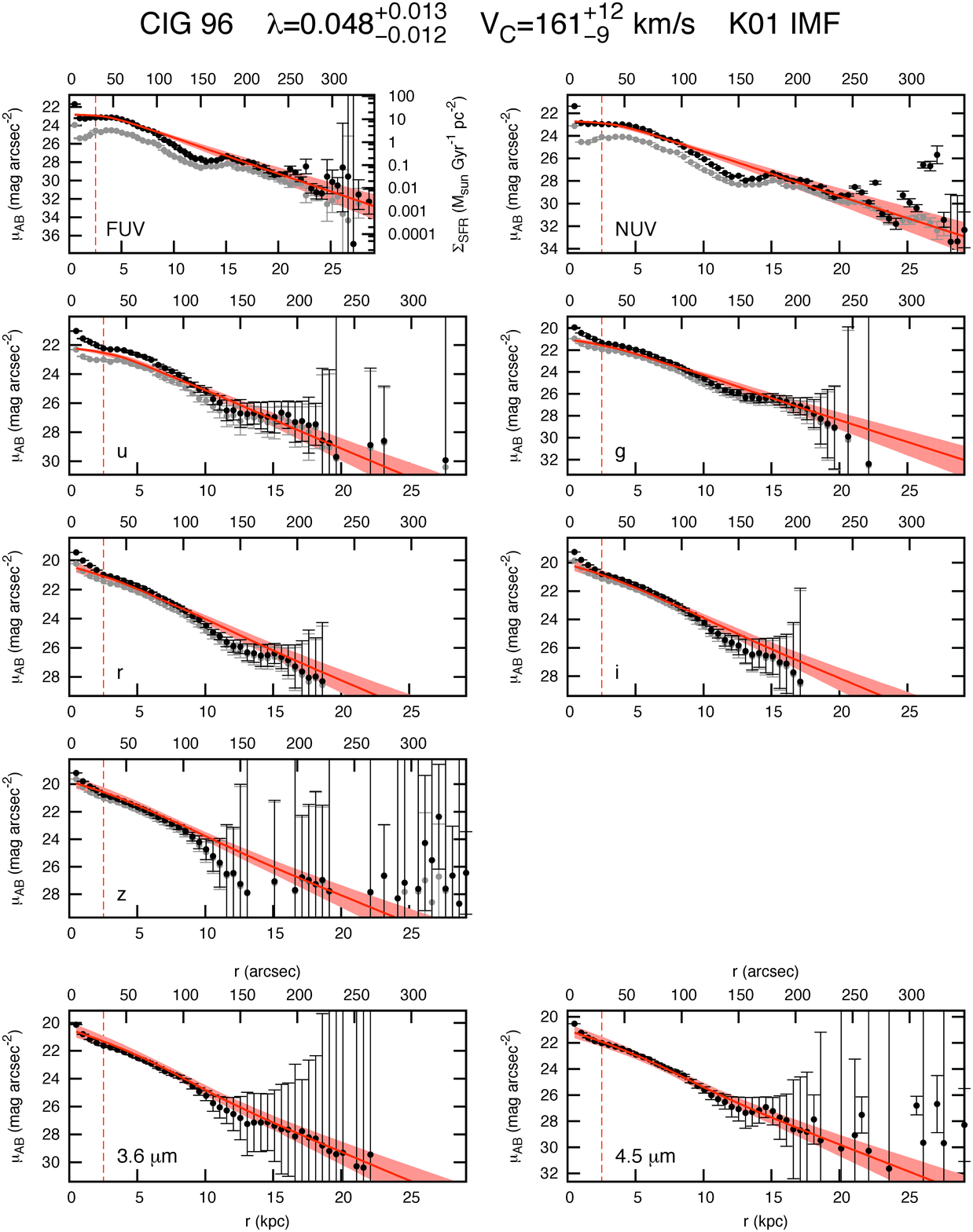}
\caption{Surface density radial profiles for the different bands: GALEX NUV and FUV, SDSS ugriz bands, and Spitzer 3.6 and 4.5~$\mu$m), in bins of 6\arcsec. The grey data points are corrected for Galactic extinction and the black data points for internal extinction as well. The latter are used for the fits using a model based on the IMF by  \citet{kroupa01} ($\lambda$ = 0.048$^{+0.013}_{-0.012}$ and $V_C$ =  161$^{+12}_{-9}$, as explained in \S~\ref{model}). The best radial profile fits are shown as (red) solid lines. The red band indicates an estimation of the error, and comprise all the fits within 2 times the minimum $\chi^2$. { The red dashed line indicates the radius we use to separate the region of the profile dominated by the bulge (left) and that by the disk (right). Only the latter region is used to perform the fit. A scale conversion for each band to $\Sigma_{\rm SFR}$ in units of M$_\odot$~yr$^{-1}$~kpc$^{-2}$ is also shown in the upper left panel. } \label{models}}
\end{figure}

We plot in each panel of Fig.~\ref{models} the de-projected surface density profile for each band considered, as well as the fits from the model. The agreement is good for all components. For UV bands, the azimuthally averaged AB magnitudes seem to be slightly underestimated in the $r$ = 10\arcsec -- 15\arcsec region, although in general they agree well. 

{
The KS law was expressed in \S~\ref{SFlaws} in the form $\Sigma_{SFR}$ $\propto$ $\Sigma_{gas}^N$, with $N$ varying
for different annuli. On the other hand, the SFR in the model is expressed as  $\Sigma_{SFR}$ $\propto$ $\Sigma_{gas}^N$ $V(R)/R$.
To check the consistency between both definitions, in Fig.~\ref{SFLaws}  (bottom) we plot the temporal evolution from (0 to 13.5~Gyr) of $\Sigma_{\rm SFR}$ and $\Sigma_{\rm gas}$ using the model, and for different radii, 5.5 (blue), 10.9 (green) and 20~kpc (red). No obvious offset is found, which indicates that this galaxy is in general following the SF law implemented in the models, which relies on average main properties of galaxies.
}

Note that the models used here are not necessarily good at reproducing the properties of galaxy disks out to
these large radii. The model itself does not include an imposed threshold, and although it is an extrapolation, it should work reasonably well  for low density regions and outskirts as long as there is no
strong threshold effects and the mode of star formation does not change. In fact these models have been tested
against observations of Low Surface Brightness galaxies in \citealt{boissier03b}.

\subsection{Predicted Rotation Curve and Gas Surface Density}

\begin{figure}
\centering
\includegraphics[width=9cm]{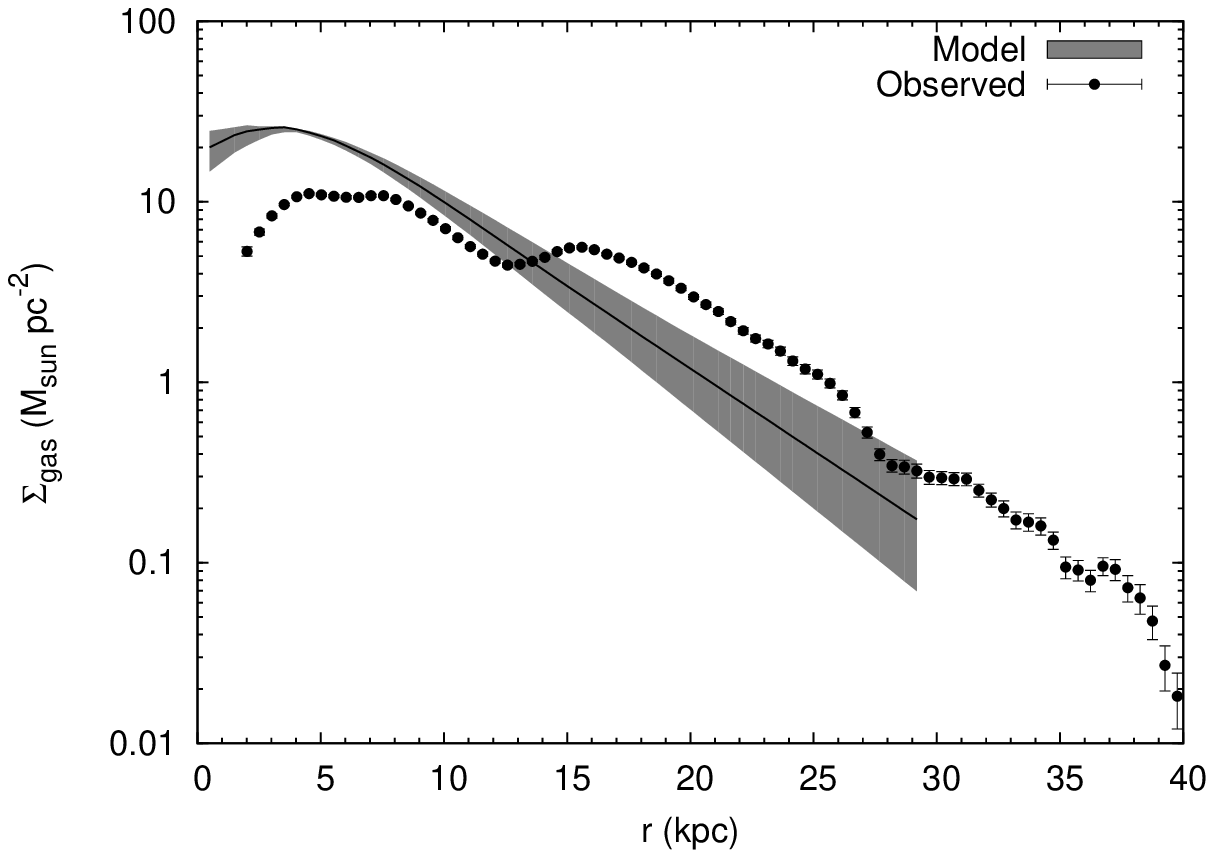}
\includegraphics[width=9cm]{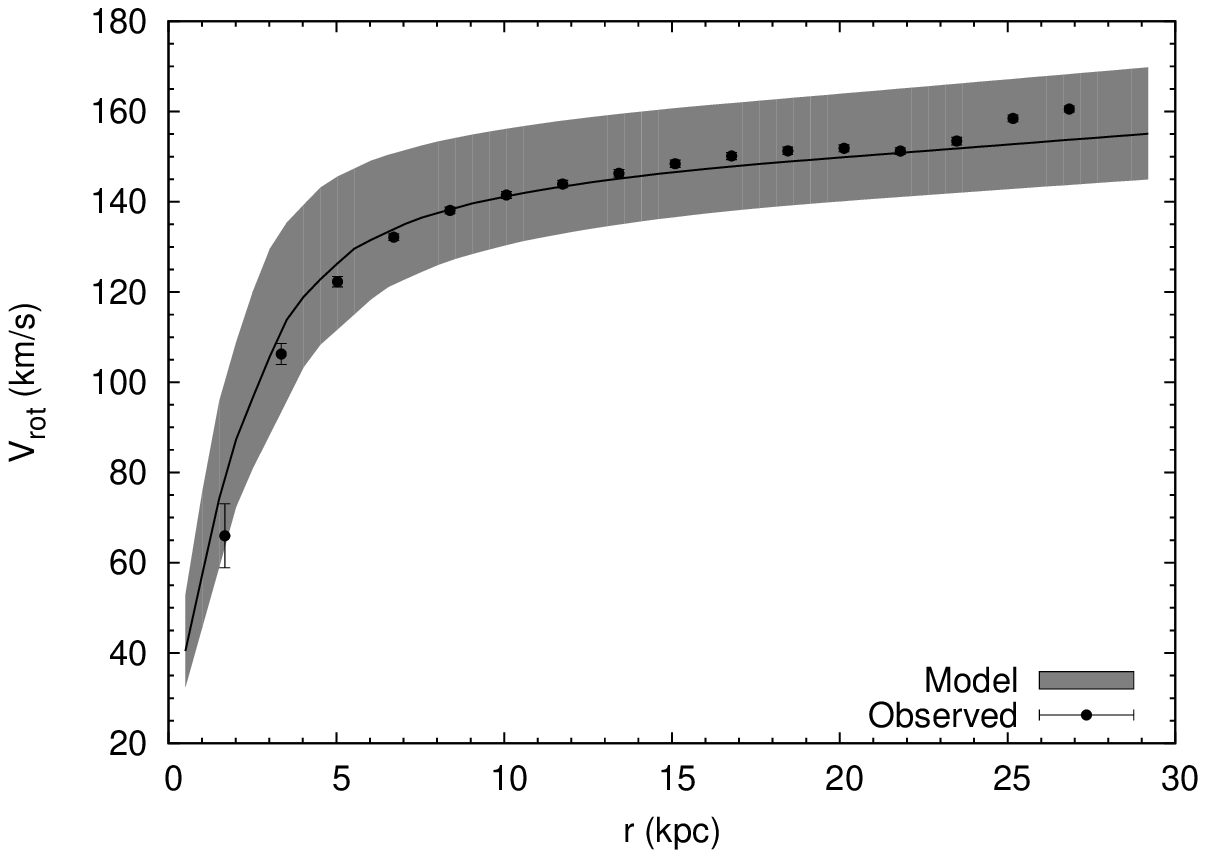}
\caption{Upper panel) Observed gas surface density ($\Sigma_{gas}$) radial profile and fit using our model  (\S~\ref{model}) based on the IMF by \citet{kroupa01}. 
Lower panel) Comparison of the observed rotation curve (average southern and northern sides, \citealt{espada05}) and that predicted by our model. 
\label{rotationcurve}}
\end{figure}

The model provides an estimation of the rotation curve and $\Sigma_{gas}$ as a function of radius. 
We compare the observed gas surface density with that predicted by the model in the upper panel of Fig.~\ref{rotationcurve}.
A small deviation is found at low radii, where most of the gaseous component is in molecular phase, and at large radii from $r$ $>$ 15~kpc ($>$ 3\arcmin) where the model underestimates the observed surface density as a result of the external pseudo-ring and the extended UV emission in general. 

{
The latter can be probably explained as a result of a past recent event that re-distributed part of the gas on time scale shorter than that probed by the different bands, because they average over the lifetime of the galaxy. UV emission, which as \hi\ should be sensitive to any perturbation, agrees relatively well to the model. This suggests that a recent event could have occurred, such as accretion, in order for the \hi\ to show this deviation with respect to the model.
}

In the lower panel of Fig.~\ref{rotationcurve} it is also seen that the agreement between the observed \hi\ rotation curve \citep{espada05} and that predicted by the model is excellent. Note that the \hi\ observed rotation curve is not used during the fit.

\section{Discussion and Conclusions} \label{conclusion}

The outer regions of disk galaxies are essential as testing sites for models of disk assembly and evolution.
XUV-disk galaxies are ideal laboratories to study low SFR and low gas surface densities conditions.
SF in the extreme (low-density,  low-metallicity) environment of the outskirts of galaxy disks can elucidate the limiting conditions for gas cloud formation and subsequent SF.  

CIG~96 is a unique object because it resides in isolation ($\gtrsim$ 3~Gyr since last major interaction) and has an extended star forming disk, even though XUV disks are usually ascribed to major interaction events. 
The observed gas surface density from 2\farcm5 to 4\farcm7 (15~kpc to 28~kpc) is larger than that expected from our model, even though the kinematics is in average well reproduced. This suggests that this extra gas and SF could be an anomaly in a (until-recently) well-behaved disk with a (non-dimensional) specific angular momentum $\lambda$=0.048$^{+0.013}_{-0.012}$ and an asymptotic velocity of the rotation curve $V_C$ = 161$^{+12}_{-9}$~\kms.

Our sensitive and high resolution data allowed us to derive the power law index of the KS law locally for different annuli, even at large radii. We find that the slope decreases with radius, being  N $\simeq$ 3.0 $\pm$ 0.3 in the inner  disk (0\farcm8 -- 1\farcm7) and decreasing to N = 1.60 $\pm$ 0.5 in the outskirts of the gaseous disk (3\farcm3 -- 4\farcm2).
{ The KS law index tends to unity as radius increases. This might reveal a privileged
relation between \hi\ and SF where the amount of molecular gas content traced by CO is small. \hi\ could thus be a good tracer of star forming gas in this low density low metallicity conditions.
More sensitive observations of both \hi\ and UV are necessary to reveal what is the limit for the KS index in extended disks.}

We found that up to the pseudo-ring feature the SFE radial profile is typical of other disk galaxies, with SFE decreasing with radius. Overall, the SFE spans typical values from $\sim$ 10$^{-9}$ yr$^{-1}$ to 10$^{-12}$ yr$^{-1}$.
Interestingly, the outer part of the \hi\ disk, the pseudo-ring feature, shows a break in the decreasing trend of the SFE presented for example by \citet{leroy08}.
The SFE for regions outside the inner disk ($r$ $>$ 1.5 $r_{25}$) is SFE $\sim$ 10$^{-11}$ yr$^{-1}$ (\S~\ref{SFlaws}). There are regions that share this constant value up to 3.5$r_{25}$, especially the southern part which  \citet{espada05} showed that it is actually a kinematically detached region to the rest of the disk. Regions with no relevant $\Sigma_{SFR}$ but above the noise level of $\Sigma_{gas}$ suggest very low SFE, of the order of about 10$^{-11}$~yr$^{-1}$.
The depletion time scales involved are quite large and thus the gas is very likely a reservoir for future SF. Even for the pseudo-ring feature that has a larger SFE for its radius, { if 1\% of the gas turns into stars in 10$^8$~yrs}, the time scale for consuming all the gas would be large, of the order of $\sim$ 10~Gyr.

Are these deviations in SFE at large radii the rule more than the exception in other XUV disks? There are two pieces of evidence that seems to show that these deviations from normal values may be commonly present in other galaxies. In Fig.~1 of \citet{leroy08}, where SFE is plotted vs galactocentric radius, the average of SFE for all the data points is larger than their fit for $r$/$r_{25}$ $>$ 1.0.
Unfortunately their calibration limit is  $r$/$r_{25}$ = 1.2.  
Probably the most relevant case to compare the SFE at large radii is that of M83. M83's extended \hi\ disk reaches, as in CIG~96, a radius $r$ $\sim$ 3.5~$r_{25}$, and it has been widely studied in the UV and \hi\ by \citet{bigiel10b}. The SFE in both cases level off at $r$ $\sim$ 1.5~$r_{25}$ with a value of 10$^{-11}$~yr$^{-1}$. Note that for M83 this corresponds to the depletion time $\tau_{dep}$ = 10$^{2}$~Gyr in Fig.~4 right panel of \citealt{{bigiel10b}}), also with a scatter 0.5 -- 1~dex.
These evidences suggest that we might have higher SFE in extended disks with respect to that extrapolated from inner radii. In order to check the generality of this result, deeper \hi\ and UV observations of more XUV disks are needed.

Next question is what mechanism is enhancing the \hi\ densities and triggering the SF in the outskirts of these galaxies?
The isolation of CIG~96 since about $\sim$ 3~Gyr ago allow us to discard the major interaction event scenario.
Thus, if gravitational instabilities induced by interactions (e.g. \citealt{gildepaz05}) are producing the observed spiral arms and pseudo-ring feature, then we can reject the major interaction event. It must be due to minor companions (hypothesis explored by \citealt{espada05}), extragalactic clouds,or dark matter sub-halos \citep{tenoriotagle81,bekki06}.
\citet{espada05} suggested the possibility of an accretion of a HI-rich companion to be the responsible of the SW kinematically detached clump and the overall outer ring-like structure.

 However, given the remarkable spiral symmetry found in the UV and \hi\ maps (i.e. two symmetric spiral arms from the N and S of the optical disk, and connecting the outer pseudo-ring feature to the E and W, respectively), one may think that it is an intrinsic property of the galaxy rather than induced by the environment. 
{ The outer spiral arms could be due to an instability or could be
driven, either by a non-axisymmetric halo \citep[e.g.][]{machado10}, or by a non-axisymmetric disk. We favor this last option,
although it is difficult to prove, or disprove since the major axis of
this oval would be more or less aligned with the kinematical major
axis. But the global morphology makes this option quite plausible. In
particular, the arms emanate symmetrically in the NW and SE, and then fall back at the other side. This is a pattern expected in the
manifold theory \citep{romerogomez06,romerogomez07,athanassoula09a,athanassoula09b,athanassoula10} if the disk is oval and the manifolds allow the orbits to make a 270$^{\rm o}$ turn, while meeting the opposite arm.
}

\acknowledgments{We thank the anonymous referee for a careful reading and very detailed report, which helped to improve this paper significantly. D.E., A.G.d.P., L.V.-M acknowledge support for this work from the GALEX Guest Investigator program under NASA grant NNX09AQ356. D.E. was supported by a Marie Curie International Fellowship within the 6$\rm ^{th}$ European Community Framework Programme (MOIF-CT-2006-40298). 
D.E., J.S. and L.V.-M were partially supported by DGI Grant AYA2008-06181-C02 and the Junta
de Andaluc\'ia (Spain)  P08-FQM-4205. 
S.V. was partially supported by a Junta de Andaluc'a Grant FQM108,
a Spanish MEC Grant AYA-2007-67625-C02-02, and a Juan de la Cierva
fellowship.
A.G.d.P and J.C.M.-M are partially financed by the Spanish Programa
Nacional de Astronom\'ia y Astrof\'isica under grants AyA2006-02358
and AyA2009-10368 and the Consolider-GTC project. A.G.d.P is also
financed by the Spanish Ram\'on y Cajal program.
{ J.C.M.-M. acknowledges finantial support from NASA JPL/Spitzer grant RSA
1374189; he also acknowledges support from the National Radio
Astronomy Observatory, which is a facility of the National Science
Foundation operated under cooperative agreement by Associated
Universities, Inc.}
GALEX is operated for NASA by the California Institute of Technology under NASA contract NAS5-98034. 
The Spitzer Space Telescope is operated by the Jet Propulsion Laboratory,
California Institute of Technology, under contract with the National Aeronautics
and Space Administration. Funding for the SDSS and SDSS-II was provided by the Alfred P. Sloan Foundation, the Participating Institutions, the National Science Foundation, the U.S. Department of Energy, the National Aeronautics and Space Administration, the Japanese Monbukagakusho, the Max Planck Society, and the Higher Education Funding Council for England. This research has made use of the NASA/IPAC
Extragalactic Database (NED), which is operated by JPL/Caltech, under contract with
NASA. We acknowledge the usage of the HyperLeda database (http://leda.univ-lyon1.fr).}

{\it Facilities:} \facility{VLA, GALEX, SDSS, Spitzer}

\begin{table}
\begin{center}
\caption{{ Local Kennicutt-Schmidt law for different annuli \label{tbl-1}}}
\begin{tabular}{cccccccc}
\hline
& & OLS(Y|X)  & & Bisector \\
Zone & {Radius range (\arcmin) / (kpc)} &  Slope (N)     &  Intercept & Slope (N)     & Intercept    \\\hline
 1 &  0.8 -- 1.7  (4.8 -- 10.2)   & 1.7 $\pm$ 0.3   & -4.0 $\pm$ 0.3     & 3.0 $\pm$ 0.3     & -5.4 $\pm$ 0.3\\
 2 &  1.7 -- 2.5  (10.2 -- 15.0)     & 1.9 $\pm$ 0.3   & -4.9$\pm$ 0.2     & 2.18 $\pm$ 0.10 & -5.07 $\pm$ 0.11\\
 3 &  2.5 -- 3.3  (15.0 -- 19.8)     & 1.4 $\pm$ 0.2   & -4.85$\pm$ 0.14 & 1.7 $\pm$ 0.2 & -5.1 $\pm$ 0.3\\
 4 &  3.3 -- 4.2  (19.8 -- 25.2)     & 1.2 $\pm$ 0.2  & -5.1$\pm$ 0.10    & 1.6 $\pm$ 0.5 & -5.3 $\pm$ 0.6\\\hline

\end{tabular}

\end{center}
\end{table}

\end{document}